\documentclass[aps,prb,reprint]{revtex4-2}
\usepackage{xcolor}  
\usepackage{placeins}
\usepackage{lineno}
\usepackage{soul}
\usepackage{graphicx,float}
\usepackage{amsmath,bm}
\usepackage{amssymb}
\usepackage{braket}
\usepackage{amsfonts}
\usepackage{dsfont}
\usepackage{comment,color}
\usepackage{lipsum}
\usepackage{hyperref}
\hypersetup{colorlinks=true,linkcolor=blue,anchorcolor=blue,citecolor=blue,filecolor=blue,urlcolor=blue,bookmarksnumbered=true,pdfview=FitB}

\begin{document}

\title{Strong metamagnetic switching and anisotropic magnetotransport in the distorted kagome magnet Tb$_6$Ti$_4$Al$_{43}$}
\author{Lipika$^{1}$}
\author{Sneh$^{1}$}
\author{Shobha Singh$^1$}
\author{Ralf Koban$^2$}
\author{Walter Schnelle$^2$}
\author{Rie Umetsu$^3$}
\author{Kaustuv Manna$^{1,*}$}
\affiliation{$^1$Department of Physics, Indian Institute of Technology Delhi, Hauz Khas, New Delhi, India 110016}
\affiliation{$^2$ Max-Planck-Institute for Chemical Physics of Solids,Nöthnitzer Straße 40
01187 Dresden, Germany}
\affiliation{$^3$Department of Physics, Institute for Materials Research, Tohoku University, Sendai, Japan}
\affiliation{$^*$kaustuvmanna@physics.iitd.ac.in}
\

\date{\today}
      
\begin{abstract}
We report a comprehensive investigation of the magnetic, electrical, and thermodynamic properties of single-crystalline Tb$_6$Ti$_4$Al$_{43}$, in which Tb atoms form a distorted kagome lattice embedded within a complex Ti–Al framework. Tb$_6$Ti$_4$Al$_{43}$ exhibits pronounced magnetic anisotropy and antiferromagnetic ordering at $\sim$  19 K. For magnetic field applied along the \textit{c}-axis (\textit{H} $\parallel$ \textit{c}), a strong metamagnetic switching is observed, accompanied by a clear asymmetric hysteresis at low temperatures, while for field applied along the \textit{ab} plane, a spin-flop transition is observed.  The effect of field-induced magnetic states is strongly reflected in magnetotransport measurements, giving rise to pronounced anomalies in magnetoresistance and an anomalous Hall contribution. A significant intrinsic contribution to the anomalous Hall conductivity is observed, highlighting the Berry curvature effects arising from the system's momentum-space topology.  Our results highlight the interplay of spin, charge, and lattice degrees of freedom in Tb$_6$Ti$_4$Al$_{43}$ and establish this distorted kagome antiferromagnet as a promising platform for field-tunable multifunctional behaviour, including electrically readable magnetic switching, spin manipulation, and magneto-mechanical functionalities.

\end{abstract}

\maketitle  
 \section{Introduction}
Kagome lattices have attracted significant attention in magnetic systems due to their unique geometry of corner-sharing triangles, which naturally introduces frustration in these systems \cite{Balents2010, Ramirez1994, kagomemetals}. Such materials have depicted many exotic, unconventional phenomena such as flat bands due to localization of electrons in real space, Dirac-like dispersions, strong geometric frustration, superconductivity, and charge density wave \cite{Kagome2025, Wang2025, Kang2020, Yin2019}. This intricate arrangement of atoms can also provide an ideal platform for realizing the Berry curvature–driven transport phenomena \cite{Berrycur,AHE}. Mn$_3$Ge and Mn$_3$Sn, non-collinear antiferromagnets, in which Mn atoms form the kagome structure, have demonstrated large anomalous Hall effects even in the absence of net magnetization, driven by symmetry-breaking spin configurations \cite{Nakatsuji2015, Nayak2016}. Similarly, Fe$_3$Sn$_2$, which is a room temperature ferromagnet, holds real space magnetic textures as well as Dirac cones and flat bands in momentum space owing to the kagome lattice frustration and anisotropy in this compound \cite{Fe3Sn2, Lipika_fe3Sn2, Li2019LargeFe3Sn2, Hou2017ObservationAnisotropy, Li2019LargeFe3Sn2}. Studies on \textit{A}V$_3$Sb$_5$ (where \textit{A} = K, Rb, Cs) series also highlight the interplay of electronic topology and correlations in kagome-based materials \cite{CsV3Sb5, Jiang2021}.
\newline
\begin{figure*}[t]
    \centering   
    \includegraphics[width=17cm]{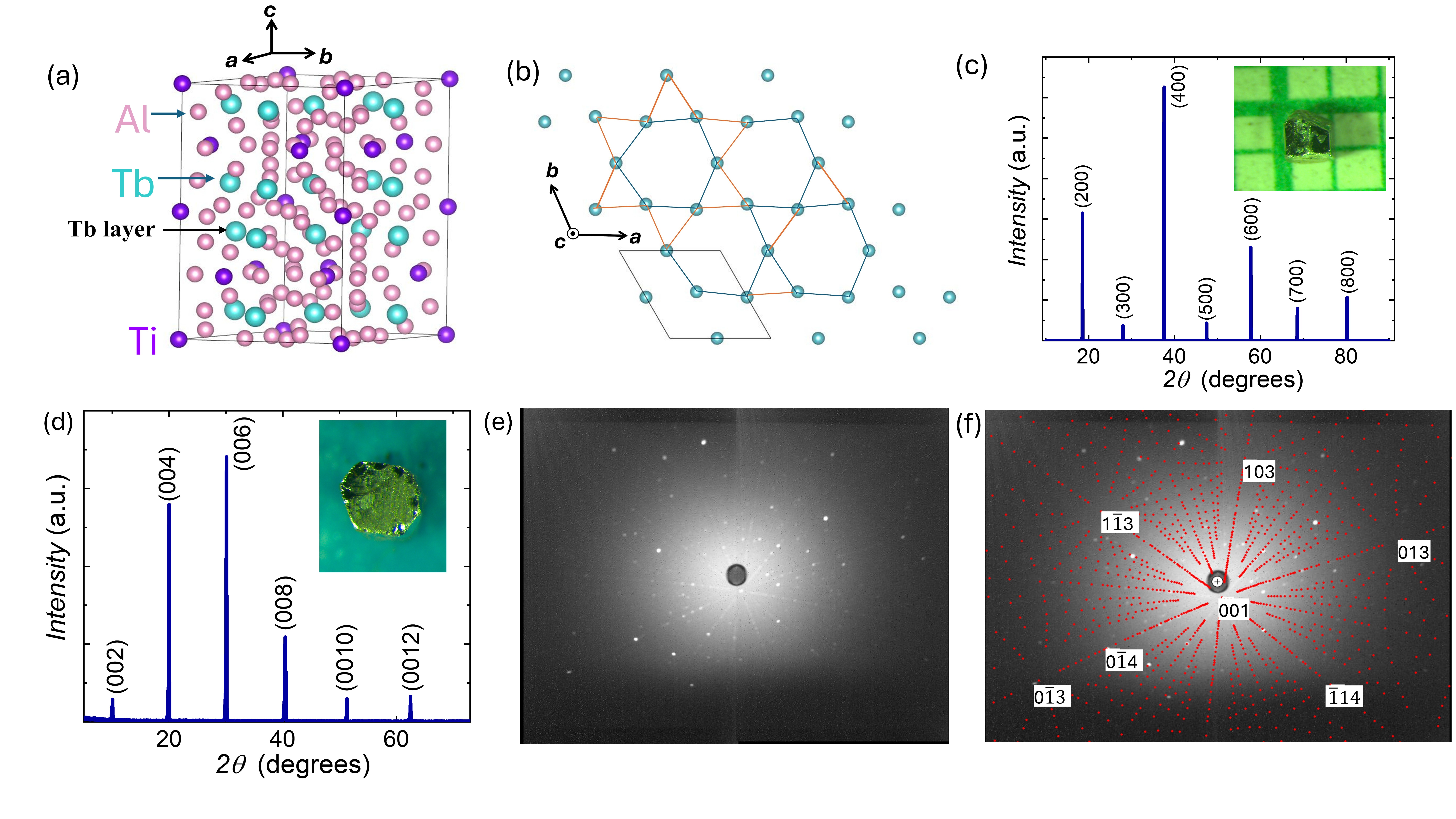}
    \caption
{(a) Atomic arrangement of Tb, Ti, and Al atoms in Tb$_6$Ti$_4$Al$_{43}$ crystal lattice. Tb, Ti and Al are shown by green, purple and pink atoms, (b) the arrangement of the Tb layer in a distorted kagome structure. XRD $\theta$ - 2$\theta$ scan for crystal orientation perpendicular to (c) (\textit{h}00) plane, (d) (00\textit{l}) plane. The corresponding crystal surface is shown in the respective insets. (e) Laue x-ray diffraction pattern for  Tb$_6$Ti$_4$Al$_{43}$ crystal, (f) simulated Laue x-ray diffraction pattern.  }
    \label{fig:my_label4}
\end{figure*}
An important characteristic of kagome material is the presence of highly degenerate magnetic ground states arising from geometric frustration. Consequently, these systems are highly sensitive to perturbations such as lattice distortions, magnetic anisotropy, and competing exchange interactions, which can lift the degeneracy and stabilize ordered, anisotropic, or field-tunable non-collinear magnetic states \cite{Balents2010, Kagome2025}. For example, in $R$$_3$Ru$_4$Al$_{12}$ ($R$ = rare earth elements), exchange interactions and magnetocrystalline anisotropy partially lift geometric frustration, allowing the system to achieve a ground state, but show field-induced transitions \cite{Gorbunov2014, Gorbunov2015, Ho3Ru4AL12}. In other systems, like HoAgGe, such distortions in the kagome lattice modifies the band structure, leading to a finite Berry curvature, resulting in an enhanced anomalous Hall effect \cite{HoAgGe}.
\newline
A similar system is Tb$_6$Ti$_4$Al$_{43}$, which hosts a distorted kagome lattice formed by magnetic Tb$^{3+}$ ions. The system retains key ingredients such as geometric frustration, strong magnetocrystalline anisotropy, and competing magnetic interactions. The magnetic properties of the broader $R$$_6$T$_4$Al$_{43}$ family, where $R$ = rare-earth elements and $T$ = Ti, V, Nb, Ta, Cr, Mo, W have been previously investigated by Wolff et al. in polycrystalline samples, where metamagnetic transitions were reported in several compounds \cite{Wol2001, Niemann1995}. In addition, single crystal growth of this family and studies on Dy$_6$Cr$_4$Al$_{43}$ reveal a strong magnetic anisotropy, indicating the importance of anisotropic effects in this family \cite{ Kangas2013, Maurya2014}.
\newline
Here, we present a comprehensive investigation of the magnetic, electrical, and specific heat properties of the single crystal of Tb$_6$Ti$_4$Al$_{43}$  to uncover the interplay between anisotropic magnetism, field-induced spin reorientation, and the charge-carrier response. It exhibits pronounced magnetic anisotropy and field-induced metamagnetic transitions, indicating the presence of competing magnetic ground states. Despite its antiferromagnetic (AFM) ground state ordering with $T_N$ $\sim$ 19 K, the system exhibits a clear anomalous Hall effect, which we attribute to a field-induced spin-polarized state emerging through low-field metamagnetic transitions for \textit{H} $\parallel$ \textit{c}. A significant intrinsic contribution to the anomalous Hall contribution suggests the presence of Berry curvature induced effects due to the momentum space topology. Furthermore, the crystal exhibits a strong magnetic anisotropy, evidenced by its anisotropic magnetic response along different crystal directions and the tendency to physically reorient itself along the magnetic easy axis (\textit{c} axis).    
\section{Experimental Details}
\begin{figure*}[t]
    \centering   
    \includegraphics[width=17cm]{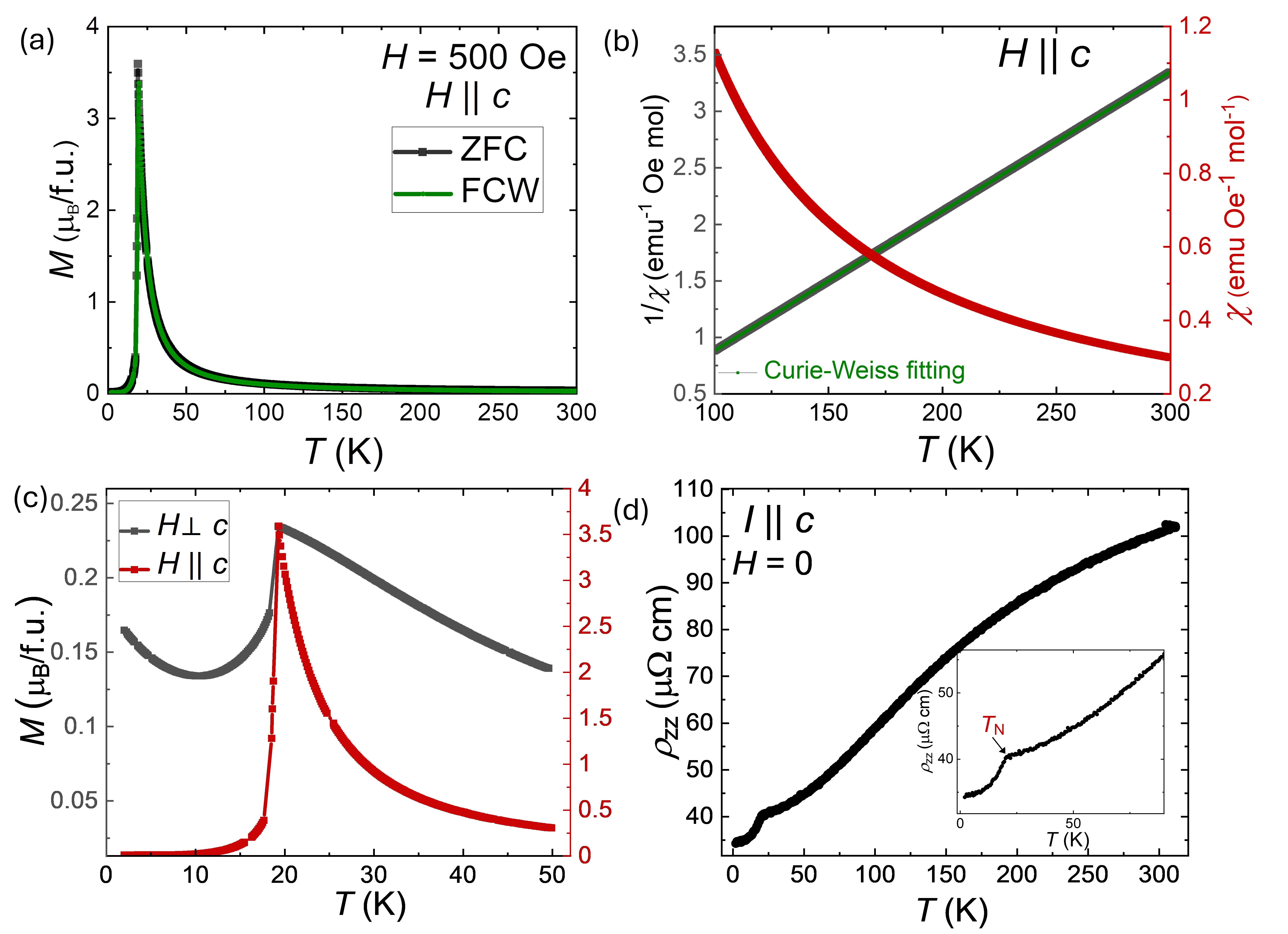}
    \caption
{(a) Temperature-dependent magnetization for zero-field cooling (ZFC) and field-cooled warming (FCW) at an applied magnetic field of 500 Oe for $H \parallel c$, (b) the Curie-Weiss fitting in paramagnetic regime from 100 K - 300 K, (c) $M(T)$ in temperature range of 2 K - 50 K for $H \parallel c$ and $H \perp c$, (d) Temperature-dependent resistivity $\rho_{zz} (T)$ at $H$ = 0, inset shows the zoomed-in resistivity with AFM transition \textit{T}$_N$.}
    \label{fig:my_label4}
\end{figure*}
Single crystals of Tb$_6$Ti$_4$Al$_{43}$ were grown using the flux method. High purity elements Tb (99.99 \%), Ti powder (Alfa Aesar, 99.99 \%), and Al (99.99 \%) were mixed in an optimized ratio of 40:3:1000 to obtain single phase Tb$_6$Ti$_4$Al$_{43}$ single crystals. All the elements were handled in an argon-filled glove box because Tb is air-sensitive. The total mass of 3 g was placed in an alumina crucible, which was then vacuum sealed in a quartz tube. The sealed ampoule was first heated to 1000 °C at the rate of 100 °C/h and held at this temperature for 24 hours to ensure a homogeneous solution. It was then slowly cooled to 750 °C at the rate of 2 °C/h. At this temperature, the ampoule was centrifuged to separate the crystals from the excess aluminium flux. The resulting crystals were silvery, with a hexagonal cross-section, and approximately 1-2 mm in size (shown in the insets of Figs. 1(c) and 1(d)). 
\newline The elemental composition of the crystal was analyzed using the Energy Dispersive X-ray spectroscopy (EDS) with Hitachi High-Technol TM 3000 with 15 keV energy. The measured elemental composition of Tb, Ti, and Al was found close to the stoichiometric ratio of 6:4:43 within the experimental uncertainty. The orientation of the crystal was confirmed using the Proto x-ray Laue diffraction with white x-rays. The $\theta$ - 2$\theta$ x-ray diffraction measurements were carried out using the HR-XRD (PANalytical Empyrean X-Ray diffractometer) with a Cu K$\alpha$ (\textit{$\lambda$} = 1.5406 \AA) monochromatic source at room temperature. Magnetic measurements, including both in-plane and out-of-plane \textit{dc} and \textit{ac} magnetization, were performed using the commercial 7 T Quantum Design Magnetic Property Measurement System (MPMS-3). A quartz sample holder was used to measure the magnetization along the \textit{c}-axis, while a brass holder was used for measurements along the hard axis of the crystal. Electrical transport and specific heat measurements were performed using the commercial Quantum Design Physical Property Measurement System (PPMS). For resistivity measurements, a standard linear four-probe geometry was employed, using platinum wire and silver paint for electrical contacts. 
\newline
\section{Results and Discussion} 
\begin{figure}[t]
    \includegraphics[width=8cm]{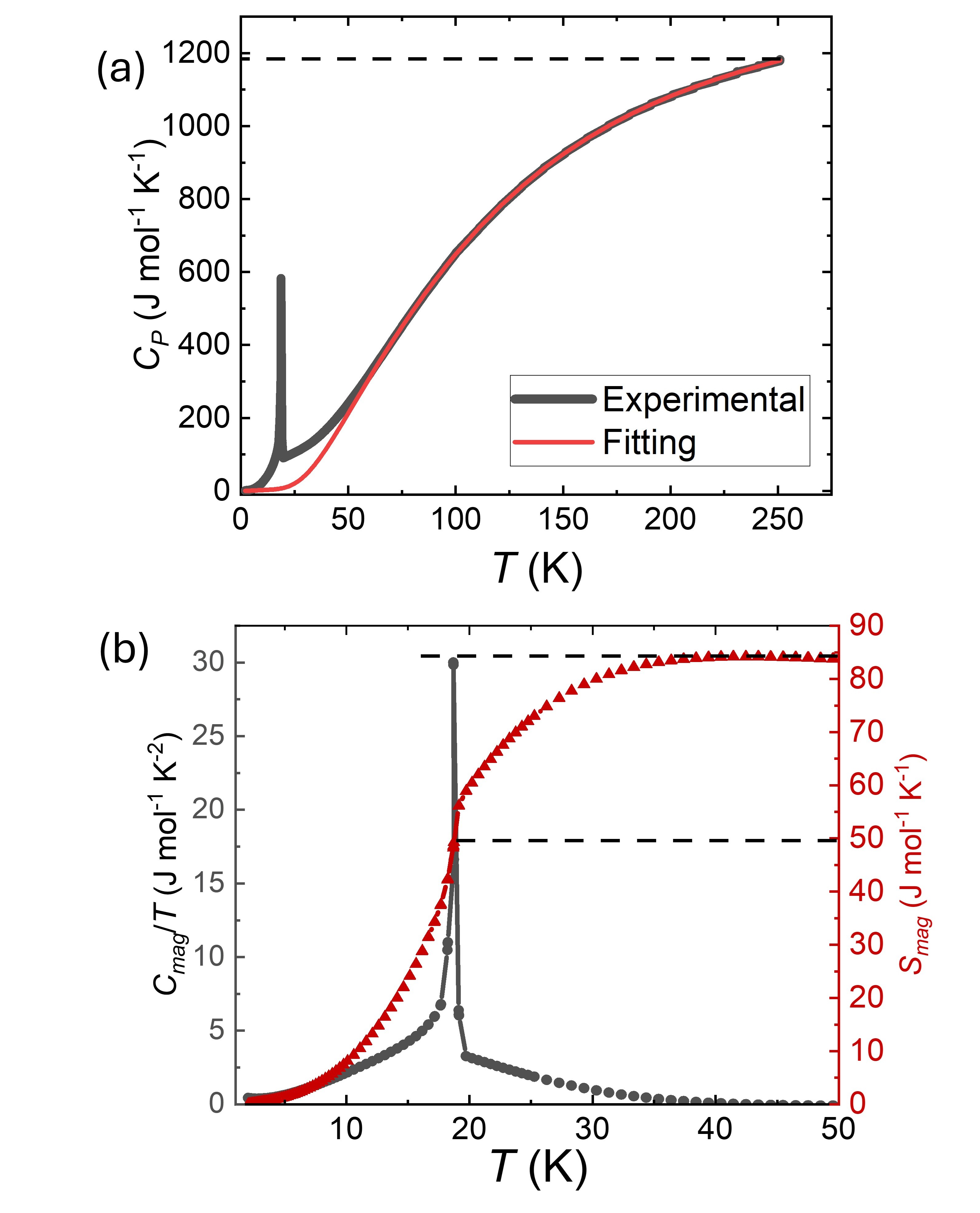}
    \caption
{(a) Black curve indicates the experimental temperature-dependent specific heat $C_p (T)$ and red curve shows the fitting of the experimental curve using the electronic and lattice contributions (fitted above 80 K), (b) temperature-dependent magnetic contribution of specific heat (C$_{mag}(T)$) and entropy (S$_{mag}(T)$) are presented by black and red curves respectively.  }
    \label{fig:my_label4}
\end{figure}
\begin{figure*}[t]
    \centering   
    \includegraphics[width=17cm]{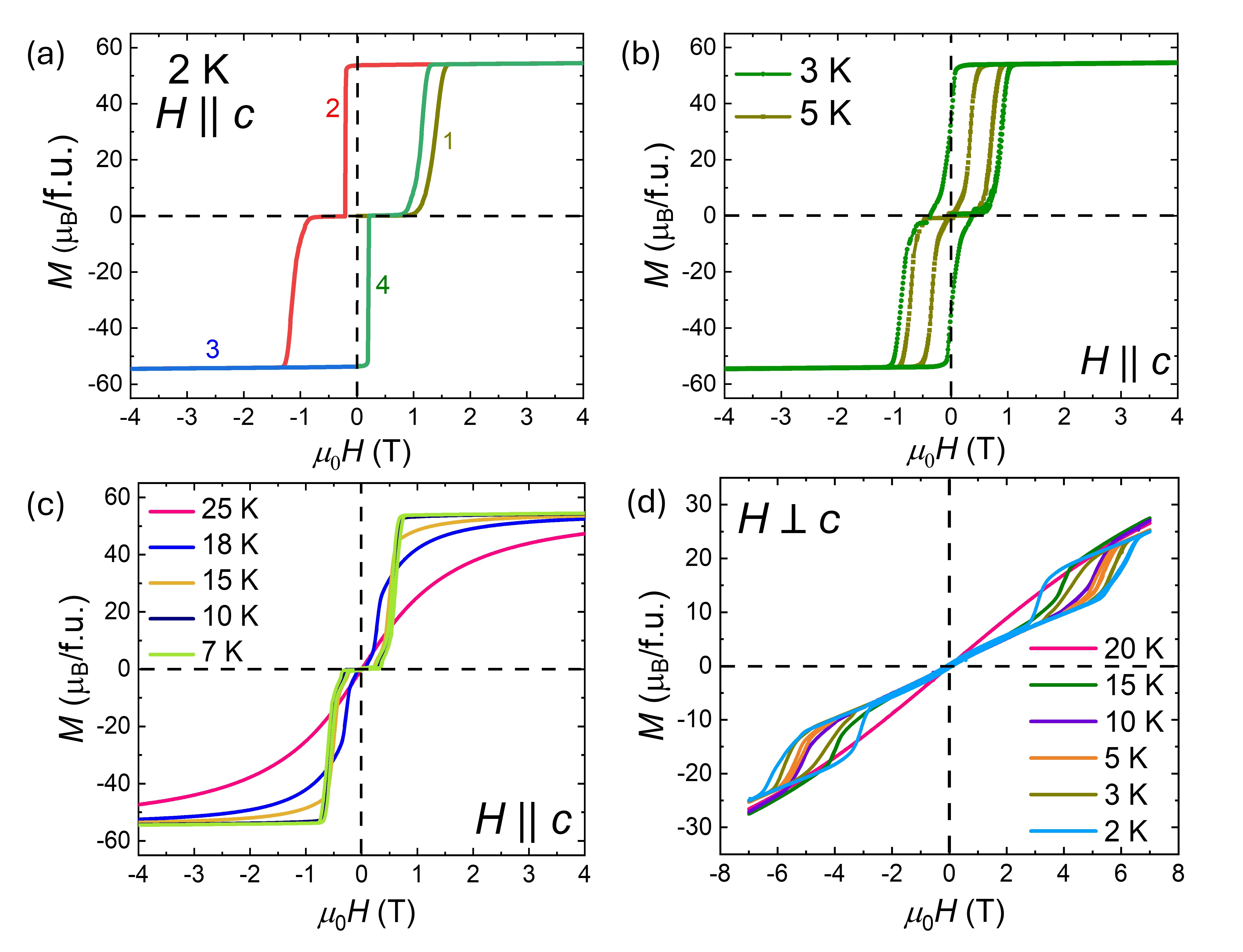}
    \caption
{Isothermal field-dependent magnetization ($M(H)$) for $H \parallel c$ at temperatures (a) 2 K, (b) 3 K and 5 K, and (c) 7 K - 25 K. (d) $M(H)$ for $H \perp c$ at different temperatures.}
    \label{fig:my_label4}
\end{figure*}
Tb$_6$Ti$_4$Al$_{43}$ crystallizes in the Ho$_6$Mo$_4$Al$_{43}$-type structure with the space group P6$_3$/mcm (No. 193) \cite{Wol2001, Karakchieva2022}. Fig. 1(a) shows the crystal structure of Tb$_6$Ti$_4$Al$_{43}$, highlighting the complex Ti-Al framework interleaved with the magnetic Tb layers. The Tb ions form a distorted kagome network within these layers (Fig. 1 (b)), providing a geometrically frustrated lattice for the magnetic interactions. The X-ray diffraction pattern is depicted in Fig. 1(c), where the peaks corresponding to the (\textit{h}00) indicates the single crystalline nature of the sample which confirms that the axis normal to the exposed crystal surface is \textit{a} axis (as shown in the Fig. 1(c) inset). Fig. 1(d) presents the x-ray diffraction $\theta$ - 2$\theta$ scan obtained, when the x-rays are incident on the hexagonal plane of the crystal (shown in the inset), indicating that the normal to the exposed surface is the \textit{c}-direction of the crystal. The lattice parameters obtained from the x-ray diffraction pattern are \textit{a} = \textit{b} = 11.03 \AA, and \textit{c} = 17.7 \AA, which is similar to the related compound Yb$_6$Ta$_4$Al$_{43}$ \cite{Wol2001}. The crystal orientation and quality are further verified by x-ray Laue diffraction. Fig. 1(e) shows the Laue diffraction pattern, the sharp and well defined spots suggest a single crystalline sample without any domains or twinning.  The simulated pattern of the Laue diffraction is indicated in Fig. 1(f), depicting the crystallographic orientation of the crystal.
\newline
To understand the magnetic behaviour of Tb$_6$Ti$_4$Al$_{43}$ single crystal, we performed temperature-dependent magnetization measurements under two conditions (i) zero-field cooling (ZFC) condition, where the sample was cooled in zero magnetic field, and the measurement was performed while warming, in the presence of 500 Oe field, and (ii) field-cooled warming condition (FCW), where the crystal was cooled in the presence of same applied field and data is collected while warming. As shown in Fig. 2(a), for \textit{H} $\parallel$ \textit{c}, the ZFC and FCW curves overlap, exhibiting a sharp peak indicative of AFM interactions with Neel's temperature (\textit{T}$_N$) $\sim$ 19 K. To further probe the nature of magnetic interactions, we performed the Curie-Weiss fitting in the range 100 K - 300 K as presented in Fig. 2(b). The extended Curie-Weiss law is given as $\chi$ = \textit{C}/(\textit{T}-$\theta$)+$\chi_o$, where \textit{C} is the Curie constant, $\theta$ is paramagnetic Curie-Weiss temperature, and $\chi_o$ is temperature independent susceptibility. The fitting yields $\theta$ $\sim$ 29 K, suggesting predominant ferromagnetic exchange interactions in the paramagnetic regime. The observation of AFM transition and positive $\theta$ indicates the competing magnetic interactions in the system. The Curie constant \textit{C} and effective moment are related as $\mu_{eff}^2$ = 8 \textit{C}. The calculated effective moment is $\sim$ 10.2 $\mu_B$ per Tb$^{3+}$, which is close to the theoretical moment  $g_J$$\sqrt{J(J+1)}$$\mu_B$ = 9.7 $\mu_B$ per Tb$^{3+}$, where \textit{J} = 6 for Tb$^{3+}$ ion.   
\newline
Temperature-dependent magnetization measured under an applied field of 500 Oe along and perpendicular to the crystallographic \textit{c} - direction in the temperature range 2 K - 50 K is depicted in Fig. 2(c). A clear magnetic anisotropy is observed for \textit{H} $\parallel$ \textit{c} and \textit{H} $\perp$ \textit{c}, indicating a strong directional-dependent magnetic response. Both orientations exhibit a sharp feature around the \textit{T$_N$}. However, for  \textit{H} $\perp$ \textit{c}, a finite magnetization persists below \textit{T$_N$}, suggesting the presence of weak uncompensated moments, which might be due to the competing interactions.
\newline
Further, to understand the effect of long-range magnetic ordering on the resistivity, we have performed temperature-dependent zero-field resistivity measurement with the current applied along the \textit{c} direction of the crystal (shown in Fig. 2(d)). The resistivity exhibits metallic behaviour, which decreases with decreasing temperature. A kink near $\sim$ \textit{T$_N$} indicates the onset of magnetic ordering, below which the drop in resistivity occurs due to the reduction of spin-disorder scattering as the long-range AFM ordering is established in the system. 
\newline
 Magnetic interactions are further probed through the temperature-dependent specific-heat measurement ${C_P}$($T$). The experimental curve is shown in black in Fig. 3(a). The specific heat at 250 K approximately reaches 1200 J mol$^{-1}$ K$^{-1}$, which is close to the 3\textit{N}R according to the Dulong Petit law, where \textit{N} is the number of atoms per formula unit and R is the Ideal gas constant. The $\lambda$ shaped peak at $\sim$ 19 K corresponds to the second-order long range AFM ordering, which is consistent with the magnetic and resistivity measurements. The \textit{C$_P$} (\textit{T}) curve is further fitted in the paramagnetic regime (above 80 K) using the electronic and lattice contributions, given by :
\begin{equation}
C(T) = \gamma T + m\, C_{\mathrm{Debye}}(T) + (1-m)\, C_{\mathrm{Einstein}}(T)
\end{equation}
where $\gamma$ is the Sommerfeld coefficient,  $\gamma$\textit{T} represents the electronic contribution, while the second and third terms correspond to Debye and Einstein phonon contributions, respectively, and $m$ is the weighting factor which accounts for the weight of the Debye and Einstein model in the lattice specific heat \cite{Specific_heat}. These terms are given by :
\begin{equation}
C_{\mathrm{Debye}}(T) = 9 N R \left(\frac{T}{\Theta_D}\right)^3
\int_0^{\Theta_D/T}
\frac{x^4 e^x}{(e^x - 1)^2} \, dx
\end{equation}
\begin{equation}
C_{\mathrm{Einstein}}(T) = 3 N R
\left(\frac{\Theta_E}{T}\right)^2
\frac{e^{\Theta_E/T}}{\left(e^{\Theta_E/T}-1\right)^2}
\end{equation}
\newline
Here $\theta$$_D$ and $\theta$$_E$ are known as Debye and Einstein temperatures, respectively. The fitted curve in the paramagnetic regime yields $m$ = 0.82, $\theta$$_D$ = 484.2 K, $\theta$$_E$ = 124.7 K, and $\gamma$ = 146 mJ mol$^{-1}$ K$^{-2}$. Using the obtained parameters, the entire temperature range is fitted as shown by the red curve in Fig. 3(a). 
\newline
The magnetic contribution \textit{C$_{mag}$} is extracted by subtracting the fitted lattice and electronic specific heat obtained using equation (1) from the experimental \textit{$C_p$ (\textit{T})} data, which is presented in Fig. 3(b). The magnetic entropy, ${S}_{mag}$, is then calculated using :
\begin{equation}
S_{\mathrm{mag}}(T) = \int_{0}^{T} \frac{C_{\mathrm{mag}}(T')}{T'} \, dT'
\end{equation}
As shown in Fig. 3(b), $S_{mag}$ reaches approximately 6Rln(3) at $T_N$, indicating that only a limited number of crystal electric field (CEF) levels are available at the ordering temperature. This suggests that the Tb$^{3+}$ multiplet is significantly split by the CEF, with only a few contributing near $T$$_N$. Even upto 50 K, the entropy does not reach the expected value of 6Rln(13), confirming that higher-lying CEF levels are thermally inaccessible within this temperature range \cite{PhysRevMaterials.3.114411, PhysRevB.82.220201}.
\newline
To probe the effect of magnetic field on the material, we performed isothermal field-dependent magnetization measurements, \textit{M}(\textit{H}), for the magnetic field applied along the \textit{c}-axis and within the \textit{ab}-plane of the crystal. For \textit{H} $\parallel$ \textit{c} at 2 K, a strongly asymmetric magnetization response is observed, as shown in Fig. 4(a). The magnetization initially increases gradually with field as expected from the AFM ground state, but around 1 T, a strong metamagnetic transition occurs which aligns moments to a saturated state (marked as curve 1). Upon decreasing the field, the magnetization remains almost constant, retaining a remanent value comparable to the saturation magnetization, and collapses to nearly zero upon application of a small reverse field (marked as curve 2). The same process is reproduced in the negative field cycle, where a sharp metamagnetic switching occurs near -1 T.
Notably, the envelope hysteresis loop lies inside the virgin magnetization curve (curve 4), indicating pronounced magnetic hysteresis and irreversible field-dependent behaviour. Such a behaviour is commonly associated with metastability and phase coexistence during field-induced magnetic transitions \cite{Manekar2001, meta, Roy2013}.
Fig. 4(b) shows the \textit{M}(\textit{H}) curves measured at 3 K and 5 K. With increasing temperature, the virgin curve progressively approaches the hysteresis loop, accompanied by a noticeable reduction in the hysteresis area, suggesting a weakening of the irreversible metamagnetic behaviour due to thermal effects. \textit{M}(\textit{H}) curves at higher temperatures are depicted in Fig. 4(c), where the metamagnetic transitions are reversible. 

\begin{figure*}[t]
    \centering   
    \includegraphics[width=17cm]{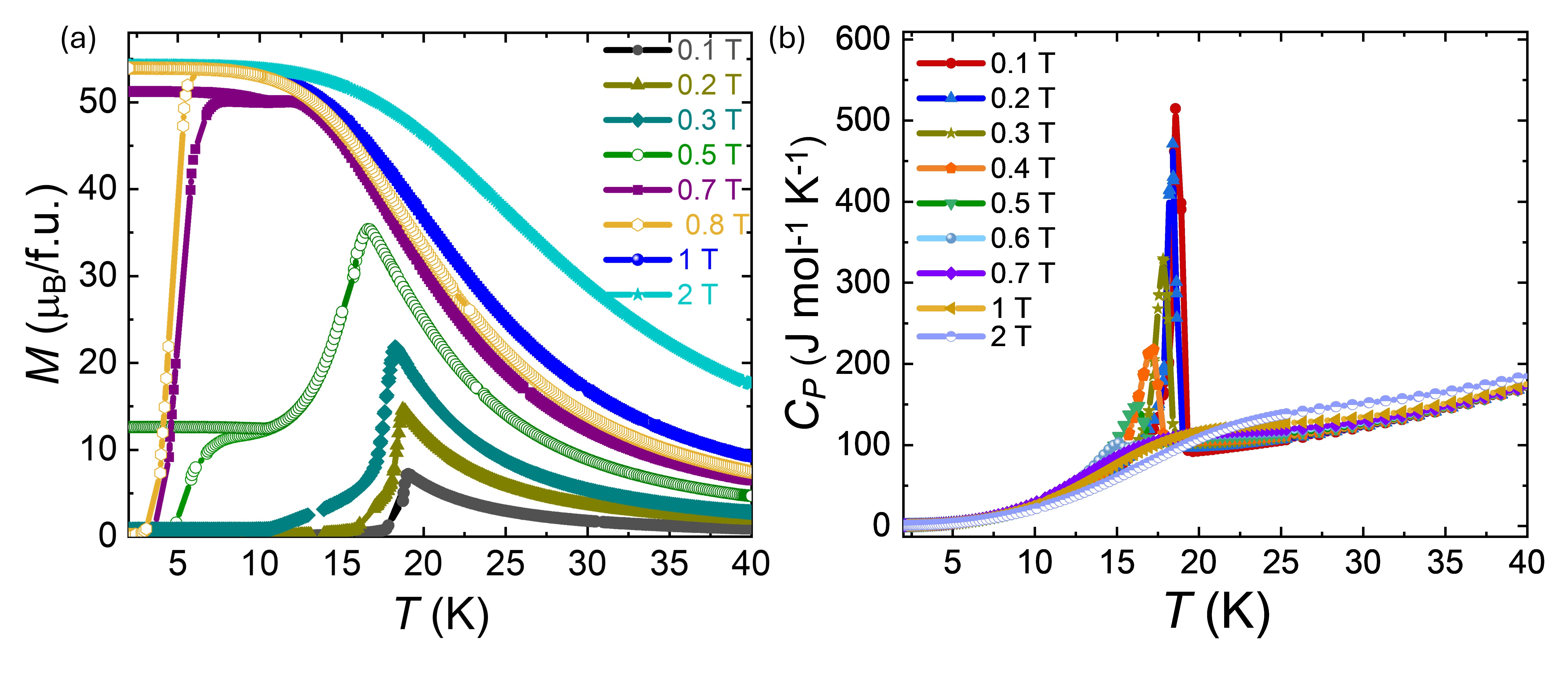}
    \caption
{(a) Temperature-dependent ZFC and FCW magnetization at different magnetic fields for $H\parallel c$, (b) temperature-dependent specific heat at different magnetic fields for $H\parallel c$. }
    \label{fig:my_label4}
\end{figure*}
Next, we measured field-dependent magnetization (\textit{M}(\textit{H})) with the magnetic field applied perpendicular to the \textit{c}-axis of the crystal (\textit{H} $\perp$ \textit{c}). The crystal was mounted on the brass sample holder, where it was firmly clamped between two small quartz rods to prevent any mechanical movement. The (\textit{M}(\textit{H})) curves depicted in Fig. 4(d) reveal irreversible spin-flop–like transitions at 2 K and 3 K, which becomes reversible at higher temperatures. Even at the highest applied field of 7 T, the magnetization does not reach saturation. This behaviour suggests a strong magnetocrystalline anisotropy in the system, where the \textit{c}-axis acts as an easy axis for the crystal. When the magnetic field is applied along the hard axis, the system undergoes a field-induced spin reorientation rather than complete alignment of magnetic moments with the field, leading to spin-flop transitions and the absence of full saturation within the measured field range.
\newline
Notably, care must be taken when mounting this crystal for measurements with the magnetic field applied along the hard axis of the crystal. Fig. S1 (see supplementary material) provides the cases of field-induced crystal reorientation observed when the field is applied within the \textit{ab} plane of the crystal (\textit{H} $\perp$ \textit{c}) \cite{SupplementalMaterial}. During magnetic and electrical measurements for \textit{H} $\perp$ \textit{c}, it was repeatedly observed that the crystal physically rotated on the sample holder, with its orientation changing so that the applied field aligned approximately along the crystal's easy axis. This suggests the presence of magnetic torque acting on the crystal, possibly due to the competition between the Zeeman energy and the magnetocrystalline anisotropy energy. When the magnetic torque becomes sufficiently large to overcome the mechanical constraint imposed by the mounting medium, the crystal undergoes field-induced physical reorientation. To obtain quantitative insights into the strength and nature of the anisotropy, torque magnetometry measurements on this crystal will be extremely useful.
\newline
\begin{figure*}
    \centering   
    \includegraphics[width=17cm]{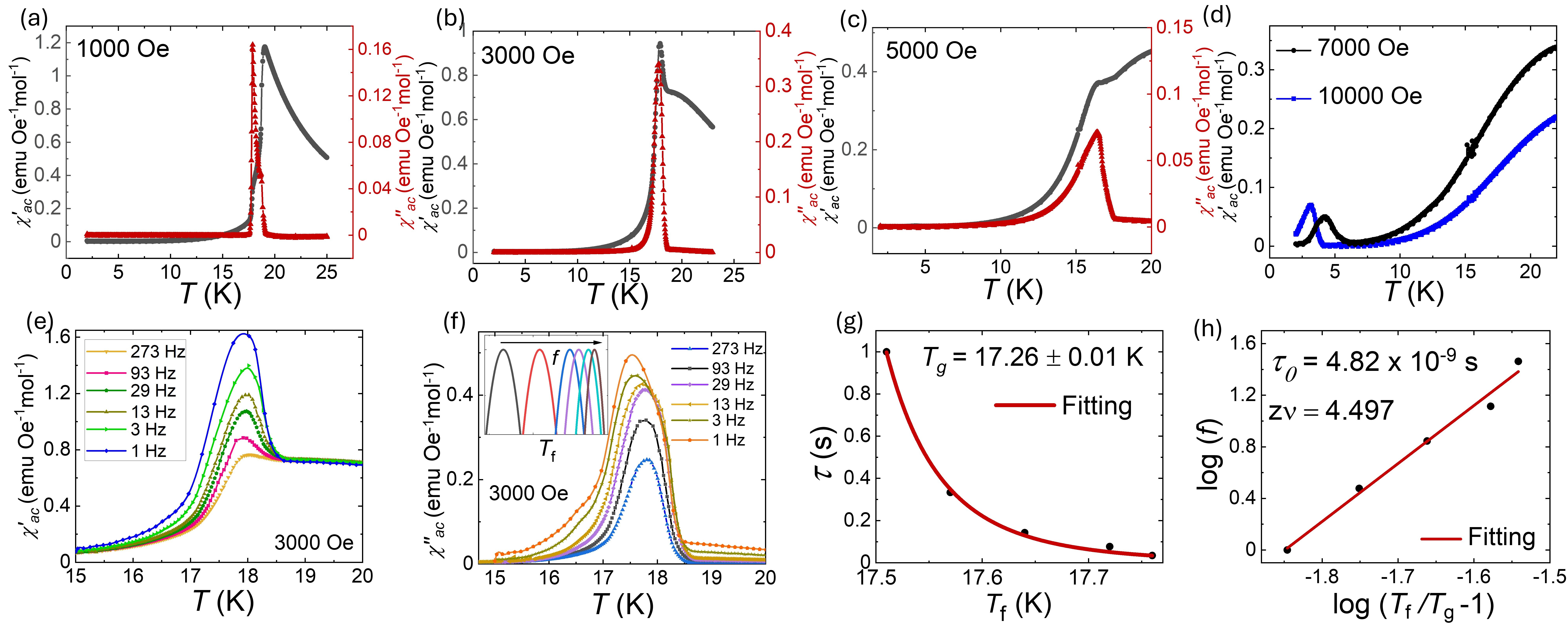}
    \caption
{Temperature dependent real and imaginary part of \textit{ac} susceptibility, $\chi\prime_{ac}$ and $\chi\prime\prime_{ac}$ corresponding to left and right \textit{y}-axis respectively, at magnetic fields (a) 1000 Oe, (b) 3000 Oe, (c) 5000 Oe. (d) $\chi\prime_{ac} (T)$ at 7000 Oe and 10000 Oe are shown in black and blue, respectively. Frequency dependent (e) $\chi\prime_{ac}(T)$, and (f) $\chi\prime\prime_{ac}(T)$ shown at 3000 Oe magnetic field. Inset of Fig. 6(f) shows the zoomed-in normalized peak shift of $\chi\prime\prime_{ac}(T)$ (g) Plot of $\tau$ vs \textit{T$_f$}, used to calculate \textit{T$_g$}. (h) Fitting of plot log(\textit{f}) vs log (\textit{T$_f$}/\textit{T$_g$} -1) and the obtained parameters. }
\label{fig:my_label4}
\end{figure*}
To further investigate the effect of the magnetic field on the magnetic transitions in this distorted kagome lattice, we performed temperature-dependent magnetization $M(T)$ measurements under zero-field-cooled (ZFC) and field-cooled warming (FCW) conditions at various applied fields (shown in Fig. 5(a)). The \textit{M}(\textit{T}) curves exhibit a clear peak corresponding to the AFM ordering temperature. For fields up to 0.3 T, the ZFC and FCW curves overlap, indicating reversibility of the AFM behaviour. However, there is also a weak kink-like feature indicating the presence of a field-induced magnetic phase near the AFM ordering temperature.
As the magnetic field increases to 0.5 T, a bifurcation between the ZFC and FCW curves emerges below ~10 K, indicating field-induced changes in the spin configuration. At 0.7 T, a sharp step-like increase in $M(T)$ emerges at $T < 10$ K, coinciding with the region of field-induced metamagnetic transition. This behaviour suggests an irreversible magnetic state where Zeeman energy effectively competes with the exchange interactions, driving a transition from the compensated AFM state to a canted or field-polarized state. In this regime, the system exhibits signatures of both AFM ordering and a ferromagnetic-like contribution at low temperatures.
With a further increase in the magnetic field, the peak corresponding to the AFM order progressively weakens, indicating suppression of the magnetic ground state. The system evolves toward a strongly canted or field-polarized magnetic state, accompanied by a pronounced bifurcation between the ZFC and FCW curves. Above 1 T, the magnetization shows a smooth temperature dependence without a distinct transition feature, consistent with a field-polarized magnetic state.
The behaviour is further verified by the temperature-dependent specific heat measurements conducted at different magnetic fields, as illustrated in Fig. 5(b). As the field increases, the peak corresponding to the transition to long-range AFM ordering shifts to lower temperatures and decreases in intensity, indicating suppression of the ground-state magnetic interaction as the system approaches the field-induced spin-saturated state.  
\newline
To further probe the spin dynamics in the system, we performed \textit{ac} susceptibility measurements, where an \textit{ac} field of amplitude 9 Oe and 93 Hz frequency was superimposed on different \textit{dc} magnetic fields. At 1000 Oe \textit{dc} field, the real part of \textit{ac} susceptibility ($\chi\prime_{ac}$) exhibits a peak at $T_N$, along with a hump like feature just below $T_N$. The imaginary part ($\chi\prime\prime_{ac}$), which captures the dissipative and dynamic magnetic response, picks up a pronounced peak corresponding to this hump, as shown in Fig 6(a). These features suggest the presence of an additional magnetic state, below $T_N$, which may be associated with spin canting or short-range correlations. This is similar to the weak kink feature observed just below the AFM transition in the presence of a small magnetic field in the $M(T)$ curve in Fig. 5(a). With an increasing \textit{dc} field to 3000 Oe, the hump in $\chi\prime_{ac}$, evolves as a well-defined peak, as also reflected in $\chi\prime\prime_{ac}$ (Fig. 6(b)), indicating stabilization of this field-induced magnetic state. This feature persists upto 5000 Oe \textit{dc} field, as presented in Fig. 6(c). At higher fields of 7000 Oe and 10000 Oe, the features corresponding to the AFM and additional magnetic state are suppressed, and a new feature appears at lower temperatures. This is attributed to a field-induced metamagnetic transition, driving the system to a spin-polarized state.
\newline
\begin{figure*}[t]
    \centering   
    \includegraphics[width = 17cm]{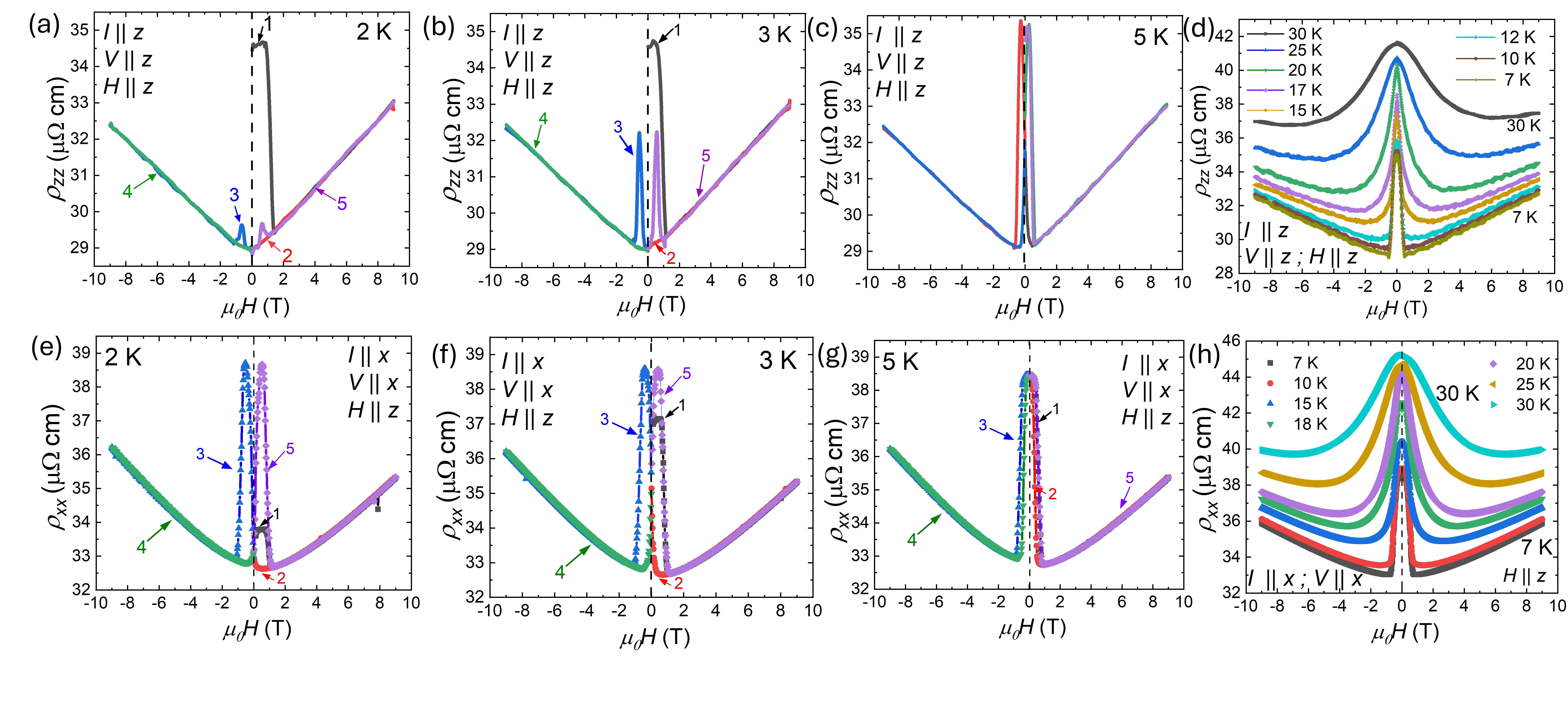}
    \caption
{Field-dependent resistivity, $\rho_{zz} (H)$ for $I \parallel z$, $H \parallel z$, and $V \parallel z$, at temperatures (a) 2 K, (b) 3 K, (c) 5 K, (d) 7 K - 30 K. Field-dependent resistivity, $\rho_{xx} (H)$ for $I \parallel x$, $H \parallel z$, and $V \parallel x$, at temperatures (e) 2 K, (f) 3 K, (g) 5 K, and (h) 7 K - 30 K.}
    \label{fig:my_label4}
\end{figure*}
To gain more insight into the field-induced magnetic state just below T$_N$, we performed frequency-dependent \textit{ac} susceptibility measurements under an applied \textit{dc} magnetic field of 3000 Oe. The \textit{ac} susceptibility shows a systematic decrease in peak intensity as frequency increases, accompanied by a discernible shift in the peak temperature. The shift is more clearly observed in the imaginary part of the \textit{ac} susceptibility. This behaviour is expected in relaxing systems; as the frequency decreases, the magnetic spins have more time to follow the \textit{ac} drive, leading to increased response in the intensity. The frequency-dependent shift in the peak suggests the presence of slow-relaxation dynamics in the system, possibly glassy behaviour. The spin relaxation time in the system is then evaluated using the critical slowing down (CSD) model, given by :
\textit{$\tau$} = \textit{$\tau$$_0$} $(\textit{T$_f$}/\textit{T$_g$} - 1)^{-z\nu}$, where \textit{$\tau$} is inverse of the applied frequency (\textit{f}), \textit{T$_f$} is the spin freezing temperature at a particular frequency \textit{f}, \textit{T$_g$} is the spin freezing temperature at \textit{f} $\rightarrow$ 0, z\textit{$\nu$} is dynamical critical constant, and \textit{$\tau$$_0$} is the microscopic spin flipping time \cite{Mydosh2014SpinGlasses,LipikaEuAuBi2026, Manna_2014, Manna2011}. The fitting of curve between $\tau$ and $T_f$ according to CSD model yields $T_g$ as depicted in Fig. 6(g). The derived T$_g$ is then used to plot $log_{10}$\textit{$f$} vs $log_{10}$(\textit{T$_f$/T$_g$} - 1), the linear fit is used to calculate the spin relaxation time, $\tau_o$ and the critical constant, presented in Fig. 6(h).  The antilog of the curve intercept (shown in Fig. 6h) corresponds to the spin relaxation time $\tau$$_0$ $\sim$ 4.82 × $10^{-9}$ s. The nanosecond-scale relaxation time is consistent with the slow collective magnetic dynamics in the system. This supports the possibility of cluster-like dynamics under the applied field, in close proximity to the AFM transition.
\newline
To understand the effect of the field-induced magnetic state on the conduction charge carriers, we performed isothermal field-dependent resistivity measurements, as presented in Fig. 7. The longitudinal resistivity, $\rho_{zz}$, measured with $I$ $\parallel$ $V$ $\parallel$ $H$ $\parallel$ \textit{z}, is presented in Figs. 7(a) - 7(d). 
At 2 K [Fig. 7(a)], the resistivity shows a weak field-dependence at low fields (indicated as 1), corresponding to the AFM ground state where the magnetization increases approximately linearly with the field. A sharp drop in resistivity is observed at the metamagnetic transition, which marks the field-induced transition from the AFM to  spin-polarized state. The suppression of spin-disorder scattering due to the alignment of moments results in a lower resistive state. At higher fields, the resistivity gradually increases, giving rise to positive magnetoresistivity. 
During the subsequent field sweep from +9 T to 0 T, the resistivity decreases smoothly without exhibiting any additional anomaly (marked as 2), indicating spin-polarized state that persists over this field range, consistent with the hysteretic magnetization behaviour shown in Fig. 4(a). Upon further field sweep towards the negative field, a clear peak is observed (marked as 3), corresponding to the metamagnetic jump in the negative cycle. This peak can be attributed to enhanced magnetic disorder, which increases the electron scattering. Upon increasing the negative magnetic field beyond the transition, the resistivity increases again on the negative side, reflecting the re-establishment of a negatively spin-polarized state with reduced spin-disorder scattering. Overall, the field-dependence of the resistivity closely follows the magnetic phase evolution, demonstrating a strong interplay between the Tb$^{3+}$ 4f moments and conduction electrons. 
Figs. 7(b) and 7(c) show the $\rho_{zz}(H)$ at 3 K and 5 K, respectively. The overall behaviour is similar to that observed at 2K. However, the intensity of the resistive peak at the metamagnetic transition increases with the temperature. This enhancement is consistent with the increased spin-disorder scattering arising from thermal fluctuations. The $\rho_{zz}(H)$ behaviour at higher temperatures is depicted in Fig. 7(d), where a similar trend is observed. Although the positive magnetoresistance corresponding to the spin-polarized state is retained, the sharp resistivity peaks are indistinguishable due to the disappearance of magnetic hysteresis in the magnetization measurements.  
\newline
\begin{figure*}[t]
    \centering   
    \includegraphics[width = 17cm]{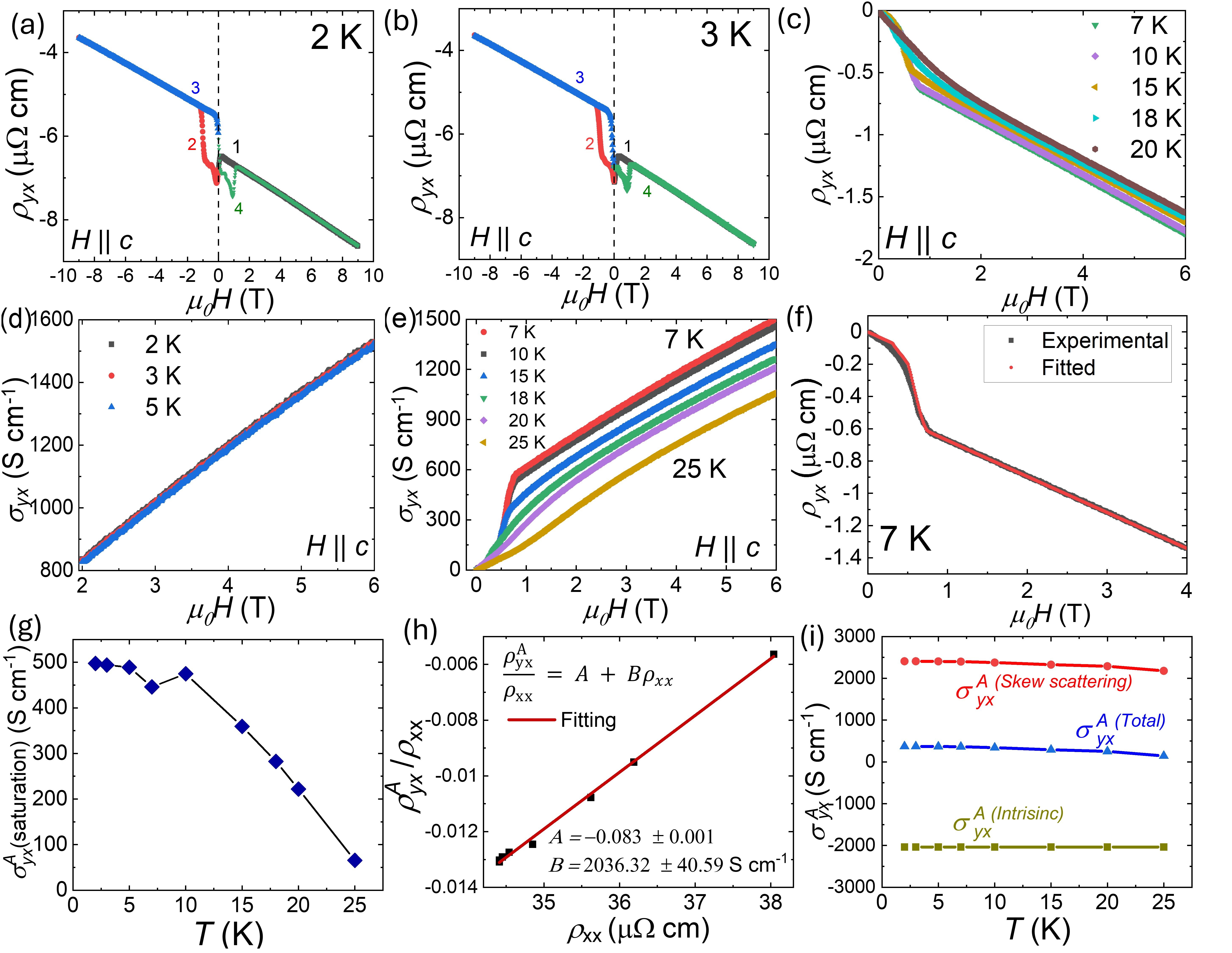}
    \caption
    {Hall response and anomalous Hall scaling in Tb$_6$Ti$_4$Al$_{43}$ for $I \parallel x$, $V \parallel y$, and $H \parallel z$. (a,b) As-measured transverse resistivity at 2 and 3 K, respectively, before antisymmetrization; numbered arrows indicate the field-sweep sequence. (c) Antisymmetrized Hall resistivity $\rho_{yx} (H)$ at selected temperatures between 7 and 20 K. (d,e) Hall conductivity $\sigma_{yx} (H)$ calculated from antisymmetrized resistivity data at 2–5 K and 7–25 K, respectively; panel (d) is restricted to the reversible high-field region. (f) Hall resistivity at 7 K fitted using $\rho_{yx}$ = $R_0 H$ + $R_SM$, accounting for ordinary and anomalous Hall contributions. (g) Temperature dependence of the saturated anomalous Hall conductivity $\sigma_{yx}^{A}$ extracted from the high-field analysis. (h) Scaling of $\rho_{yx}^{A}$/$\rho_{xx}$ with $\rho_{xx}$; the red line represents a fit to $\rho_{yx}^{A}$/$\rho_{xx}$ = A + B $\rho_{xx}$, with fitted parameters and uncertainties indicated. (i) Temperature dependence of the anomalous Hall conductivity reconstructed from the scaling fit and its contributions assigned to skew scattering and the mostly dominated intrinsic mechanisms.}
    \label{fig:my_label4}
\end{figure*}
The transverse magnetoresistivity, measured with $H \parallel z$ and $ I \parallel V \parallel x$ is presented in Figs. 7(e) - 7(h). The overall resistivity behaviour is qualitatively similar to the longitudinal configuration, confirming that the transport anomalies originate from the same field-induced magnetic transitions. Compared to the longitudinal geometry, the peak anomalies are considerably more pronounced in this configuration, indicating that the in-plane magnetic transport is more sensitive to the reorientation of the Tb moments. 
Attempts were also made to measure the resistivity with magnetic field applied in the \textit{ab} plane. However, reliable measurements could not be performed in that configuration as the sample exhibited a tendency of rotation, eventually leading to the failure of electrical contacts.
\newline
To further investigate the influence of magnetic states on the charge carriers, Hall measurements were performed. The field-dependent Hall resistivity ($\rho_{yx}$($H$)) of Tb$_6$Ti$_4$Al$_{43}$ was measured at various temperatures with $H \parallel z$, $V \parallel y$, and $I \parallel x$, as depicted in Fig. 8. At low temperatures, the $\rho_{yx}(H)$  (Fig. 8(a)) shows an irreversible Hall response, closely following the hysteretic magnetization behaviour associated with the field-induced metamagnetic transition. During the initial field sweep, the Hall resistivity changes abruptly across the metamagnetic transition and subsequently becomes reversible once the system enters the field-polarized state. Since the Hall signal is irreversible below 5 K, the conventional antisymmetrization procedure 
\begin{equation}
\rho_{yx} = \frac{\rho_{yx}(\mu_{0}H) - \rho_{yx}(-\mu_{0}H)}{2}
\end{equation}
cannot be applied over the entire field range, as it would artificially suppress the intrinsic hysteretic Hall response. Therefore, the Hall resistivity data are presented in their measured form at low temperatures in Figs. 8(a) and 8(b). In the high field region, where the Hall signal is reversible, the antisymmetrization procedure can be employed to minimize the effects due to the longitudinal resistivity component that might be contributing due to the slight voltage-contact misalignment. Fig 8(c) presents the antisymmetrized $\rho_{yx}$($H$) from 7 K to 20 K.
The corresponding Hall conductivity is obtained using $\sigma_{yx}$ = -$\rho_{yx}$/($\rho_{yx}^2$+$\rho_{xx}^2$). For 2 K - 5 K, the antisymmetrized Hall data in the reversible region above 2 T are utilized to obtain the $\sigma_{yx}$ as shown in Fig. 8(d).
For temperatures 7 K - 25 K, the resulting Hall conductivity in the entire field range, $\sigma_{yx}$, is presented in Fig. 8(e). 
\newline
The experimentally measured $\rho_{yx}(H)$ can be expressed as the sum of the ordinary Hall effect (OHE) and the anomalous Hall effect (AHE),
\begin{equation}
\textit{$\rho_{yx}$} =\textit{$R_{0}\mu_{0}H$}+ \textit{$R_{s}M$}
\end{equation}
where \textit{R$_0$} is the ordinary Hall coefficient, \textit{R$_s$} is the anomalous Hall coefficient, and \textit{M} is the isothermal magnetization \cite{Berger1970, AHE, karplus}. The experimental and fitted curves at 7 K are presented in Fig. 8(f), indicating the magnetic origin of the anomalous Hall effect in Tb$_6$Ti$_4$Al{$_{43}$.  
 \newline
 Fig. 8(g) depicts the anomalous Hall conductivity (AHC), which is derived by subtracting the ordinary Hall contribution, determined from the linear fit of the high-field region of the $\sigma_{yx}(H)$ (Figs. 8(d) and 8(e)) from the total Hall conductivity. At 2 K, the AHC ($\sigma_{yx}^{A}$) reaches 500 S cm$^{-1}$, and gradually decreases in magnitude with increasing temperature, following the suppression of magnetic field-induced transitions. 
 To distinguish the different contributions to the AHE, the scaling relation was employed \cite{karplus, Berger1970, Onoda}.
\begin{equation}
\rho_{yx}^{A} = A\rho_{xx} + B\rho_{xx}^{2}
\end{equation}
Here, the linear term, \textit{A}$\rho_{xx}$, is associated with the extrinsic skew scattering mechanism, whereas the quadratic term, \textit{B}$\rho_{xx}^{2}$, contains contributions from both the intrinsic Berry curvature and the extrinsic side-jump mechanism. 
Fig. 8(h) presents the plot of {$\rho_{yx}^{A}$/$\rho_{xx}$} versus {$\rho_{xx}$}, from which the coefficients \textit{A} and \textit{B} are obtained by fitting to the scaling relation (7).  The corresponding temperature dependence of AHC arising from the skew scattering and the quadratic (intrinsic + side jump) contributions is shown in Fig. 8(i). The quadratic contribution reaches a magnitude of $\approx$ 2000 S cm$^{-1}$, indicating a substantial contribution from the intrinsic and side jump mechanism. The scaling relation cannot separate the contributions from each of the mechanisms. The possible magnitude of the side jump mechanism can be theoretically estimated using $\sigma_{yx}^{sj}$ $\sim$ $(e^2/ha)$($\epsilon_{so}$/E$_F$), where $a$ is the lattice constant, $\epsilon_{so}$ is the spin-orbit interaction energy, and $E_F$ is the Fermi energy. For a conventional metallic system, ($\epsilon_{so}$/E$_F$) is of the order of 10$^{-2}$ or smaller, and hence the estimated side jump for Tb$_6$Ti$_4$Al$_{43}$ contributes around $\sim$ 3.5 S cm$^{-1}$ for \textit{a} $\sim$ 11 \AA  \cite{Onoda, Nozires1973, Liu2018, Xu2026Fe3Ge}. This is substantially smaller than the experimentally obtained quadratic contribution magnitude of $\sim$ 2000 S cm$^{-1}$. The large quadratic contribution is therefore consistent with an intrinsic mechanism generated by the topological band structure. In contrast, the contribution of skew scattering reaches $\approx$ 2500 S cm$^{-1}$. The skew-scattering and quadratic contributions have opposite signs and largely compensate each other, resulting in a net anomalous Hall conductivity of approximately 500 S cm$^{-1}$. Although skew scattering constitutes an extrinsic contribution, the large contribution due to the intrinsic AHC is particularly interesting, as it reflects the Berry curvature of the occupied electronic states and may be associated with the underlying electronic band structure of Tb$_6$Ti$_4$Al$_{43}$ \cite{AHE, Berger1970, Onoda}.
\newline
\section{Conclusion} 
{
 In summary, we have synthesized single crystals of the distorted kagome compound Tb$_6$Ti$_4$Al$_{43}$ and demonstrated the coupling of its electrical, magnetic and thermodynamic responses.  The material orders antiferromagnetically around 19 K and shows pronounced magnetic anisotropy, likely due to the distorted Tb kagome magnetic layer. For $ H \parallel c$, a sharp field-induced metamagnetic transition at an applied field of 1 T, switches the system from the AFM to a  spin-polarized state which collapses upon application of a small negative field. This is reminiscent of the unequal write and erase fields used in magnetic switching. For a field applied along the hard axis, the crystal exhibits spin-flop behaviour and a tendency to rotate towards the easy axis.
The field-induced magnetic transitions are also reflected in the electrical transport measurements. Both longitudinal and transverse resistivity follow the magnetization, leading to jumps at the metamagnetic switching. This shows the strong coupling between the 4f electrons and the conduction carriers in the system, giving a direct electrical readout of the magnetic state. Despite being AFM, the anomalous Hall contribution of 500 S cm$^{-1}$ is observed corresponding to the field-induced transitions. A large contribution to AHC arises due to the intrinsic mechanisms, suggesting the presence of topological electronic band structure.
The combination of strong magnetic anisotropy, field-controlled switching between the AFM and spin-polarized states, field-induced mechanical reorientation, and intrinsic anomalous Hall contribution demonstrates the interplay between spin, charge, and topology in Tb$_6$Ti$_4$Al$_{43}$. This work suggests that the distorted kagome $R_6T_4$Al$_{43}$ family can provide a fertile platform for exploring electrically readable magnetic switching, spin manipulation, and magneto-mechanical functionalities.

\begin{acknowledgments}
We acknowledge funding support from the Max Planck Society under Max Planck-India partner group project and Science and Engineering Research Board, DST, Government of India, via grant no: CRG/2022/001826. Lipika would like to thank Prime Minister Research Fellowship grant (PMRF Id : 1402122)  for the research support and fellowship. We acknowledge ICC-IMR, Tohoku University for the research support (Proposal no. : 202412-CRKKE-0503). Authors also thank Central research facility and Nanoscale research facility at IIT Delhi for providing the materials characterization facilities, like PPMS, MPMS, EDX, XRD etc.
\end{acknowledgments} 
\bibliography{bibliography.bib}
\end{document}